\documentstyle[12pt]{article}
\newcommand{\vj}{\mbox{\boldmath $j$}}
\newcommand{\vD}{\mbox{\boldmath $D$}}
\newcommand{\vp}{\mbox{\boldmath $p$}}

\newcommand{\Pitild}{\mbox{${\tilde{\Pi}}$}}
\newcommand{\vPitild}{\mbox{${\tilde{\boldmath \Pi}}$}}

\newcommand{\epsi}{\mbox{$\varepsilon$}}

\title{\bf  
Dirac-Fock Current and Magnetic-Moment \\
in the Relativistic Approach \\}

\author{ \\ Tomoyuki Maruyama\\  \\
Advanced Science Research Center, \\
Japan Atomic Energy Research Institute,\\
Tokai, Ibaraki 319-11, Japan }
\date{}

\begin{document}

\maketitle

\begin{abstract}

We define the Dirac-Fock current to hold the current conservation 
in the momentum-dependent Dirac fields, and discuss its effect 
on the magnetic-moment in the nuclear matter.
It is shown that the Dirac-Fock current can largely reduce 
the magnetic-moment even in the isovector case, 
whose value has been too big due to the small effective mass 
in the usual relativistic Hartree approximation.

\end{abstract}

\vfil
\eject
\newpage

The past decades have seen many successes in the relativistic treatment of
the nuclear many-body problem.
The relativistic framework has big advantages in several aspects \cite{Serot}:
a useful Dirac phenomenology for the description of nucleon-nucleus
scattering \cite{Hama}, the natural incorporation of the spin-orbit force 
\cite{Serot} and the saturation properties in the microscopic treatment
with the Dirac Brueckner Hartree-Fock (DBHF) approach \cite{DBHF}. 

These results conclude that there are large attractive scalar and 
repulsive vector-fields, and that the nucleon effective mass
is very small in the medium.
However this small effective mass makes some troubles in the nuclear
properties: too big magnetic-moment \cite{magmom} and 
too big excitation energy of
the isoscalar giant quadrupole resonance (ISGQR) state \cite{Nishi}. 
As for the isoscalar magnetic-moment, this enhancement is canceled 
by the Ring-Diagram contribution \cite{Ring};
this relation is completely realized by the gauge invariance \cite{Bentz}. 
As for the isovector one, however, this contribution does not play a strong
role because the symmetry force is not efficiently large.

In this subject, by the way, most of people believe that 
the momentum-dependence of the Dirac fields is negligible 
in the low energy region, particularly below the Fermi level.
A momentum-dependence of the Schr\"odinger equivalent potential
automatically emerges as a consequence of the Lorentz transformation properties
of the vector-fields without any explicit momentum-dependence of the scalar and
vector fields.
In fact, only very small momentum-dependence has appeared in 
the relativistic Hartree-Fock calculation \cite{RHF}.

In the high energy region, however, the vector-fields must become very small 
to explain the optical potential of the proton-nucleus elastic scattering 
\cite{Hama,KLW1}, and the transverse flow in the heavy-ion collisions 
\cite{TOMO1}.
This fact suggests us that the momentum-dependent part is not actually small
though it has not been clearly seen in the low energy phenomena.
After this consideration we naturally have a question whether
the momentum-dependence of the Dirac fields is really not important.
Since the momentum-dependent fields break the current conservation, 
furthermore, we have to define the new current caused by the vertex correction.

In this paper, thus, we  define a new current to hold the 
current conservation in the momentum-dependent Dirac fields, 
and discuss its effect on the magnetic-moment.

Let us consider the infinite nuclear matter system.
The nucleon propagator in the self-energy $\Sigma$ is 
given by 
\begin{equation} 
S^{-1}(p) = \gamma p - M - \Sigma(p),
\label{prop}
\end{equation} 
where $\Sigma(p)$ has a Lorentz scalar part $U_s$ and a Lorentz vector 
part $U_{\mu}(p)$ as
\begin{equation}
\Sigma(p) = - U_{s}(p) + \gamma^{\mu} U_{\mu}(p).
\end{equation} 
For the future convenience we define the effective mass and the kinetic 
momentum as
\begin{eqnarray}
M^{*}(p) & = & M - U_{s}(p) ,
\nonumber \\
\Pi_{\mu}(p) & = & p - U_{\mu}(p) .
\end{eqnarray}
The single particle energy with momentum {\vp} is obtained as
\begin{eqnarray}
\varepsilon({\vp}) & = & p_0 |_{on-mass-shell}
\nonumber \\
& = & \sqrt{ \Pi^2({\vp}) + M^{*2} } + U_{0}(\vp) . 
\end{eqnarray}
Then the detailed form of the nucleon propagator eq. (\ref{prop}) is
represented by
\begin{equation}
S(p) = \{ \not{\Pi}(p) + M^{*}(p) \} 
\{ \frac{1}{\Pi^2 - M^{*2} + i\delta } - 
2 i \pi n(\vp) \delta ( {\Pi}^2 - M^{*2} ) \}
\label{propd}
\end{equation}
with the momentum distribution $n(\vp)$.
 
If the self-energy has a momentum-dependence, the current operator
must be also changed.
We then define the current vertex $\Gamma^{\mu}(p+q,p)$ as  
\begin{equation}
\Gamma^{\mu}(p+q,p) = \gamma^\mu + \Lambda^\mu(p+q,p).
\end{equation}
From the Ward-Takahashi identity
the density-dependent vertex correction $\Lambda^\mu$  must be satisfied with
\begin{equation}
q_\mu \Lambda^{\mu}(p+q,p) = - \Sigma(p+q) + \Sigma(p).
\end{equation}

We call the new additional current 
${\hat j} = {\bar \psi} \Lambda^\mu \psi$ 
the Dirac-Fock current, which is produced by the momentum-dependence of 
the Dirac fields, because the momentum-dependent part is manily caused by the
Fock-diagram.
Next we discuss a role of this current to the magnetic-moment.

In the limit $q \rightarrow 0$ this vertex correction becomes
\begin{equation}
\Lambda^{\mu}(p) = \lim_{q \rightarrow 0} \Lambda^{\mu} (p+q,p)
= - \frac{\partial}{\partial p_\mu} \Sigma (p) .
\end{equation}
Using the above vertex correction, the current density of the whole system
is given as
\begin{eqnarray}
\vj_\mu & = & \int \frac{d^4 p}{(2 \pi)^4} {\rm Tr} \{ \Gamma_\mu S(p) \}
\nonumber \\
 & = & \int \frac{d^3 p}{(2 \pi)^3} n({\bf p}) 
\frac{\Pitild_{\mu}(p)}{\Pitild_{0}(p)} ,
\end{eqnarray}
where $\Pitild_{\mu}$ is defined by
\begin{eqnarray}
\Pitild_{\mu} (p) & = & \frac{\partial}{\partial p^\mu} V(p)
\nonumber \\
 & = & 
\frac{1}{2} \frac{\partial}{\partial p^\mu} \{ \Pi^{2}(p) - \ M^{*2}(p) \} .
\end{eqnarray}

Let us consider the one-particle state on the Fermi surface.
The nucleon space current can be written as
\begin{equation}
\vj = \frac{{\bf \vPitild}(p)}{\Pitild_{0}(p)} |_{|{\vp}| = p_F}
= \vD_{\vp} \varepsilon ({\vp}) |_{|{\vp}| = p_F},
\end{equation}
where the total derivative ${\vD}_{\bf p}$ is defined on the on-mass-shell
condition: $p_0 = \epsi ({\vp})$.
The above equation completely agrees with that derived by the semi-classical
way \cite{KLW1}.

In the non-relativistic framework the effective mass  is defined by
\begin{equation}
M^{*}_{L} = 
(2 \frac{d}{d \vp^2} \epsi({\vp}) )^{-1}
|_{|{\vp}| = p_{F}},
\end{equation}
which is so called the '{\bf Landau mass}'.
Then the above spacial current is
\begin{equation}
{\vj} = \frac{{\vp}_F}{M^{*}_{L}}.
\end{equation}
In our case including the momentum-dependent Dirac fields, the value of 
the Landau mass $M^{*}_{L}$ cannot be uniquely determined from 
the relativistic effective mass $M^{*}$ while in the Hartree approximation 
the Landau mass becomes $M^{*}_{L} = \Pi_0({\bf p}_F)$.

Now the main problem is how to determine the value of the Landau mass 
$M^{*}_{L}$.
In the theoretical aspect it should be derived by making use of DBHF 
with the momentum-dependent self-energies.
However it is not so easy technically.
Recently Huber et al. \cite{DBHF2} have introduced the momentum-dependence 
into the DBHF calculation, but they have not given this value.
As discussed later, very small contribution of 
the momentum-dependence largely changes the value of $M^{*}_{L}$ even if
it does not change the saturation property.
Hence the ambiguity involved in their calculation, for example, 
the basic nuclear force, is not as small as to fix this value.

Hence we need to determine the value of $M^{*}_{L}$ from 
the experimental analysis.
One candidate is the analysis of $p-A$ elastic scattering data, 
but this method cannot be used in very low energy region near the Fermi level.
In other observables the excitation energy of ISGQR must be most sensitive
to the Landau mass because this energy is almost determined
by the mean kinetic energy, and the surface effect is not significant.
In addition, the relativistic approach does not change its microscopic picture
\cite{Nishi}; the strength of the resorting force is determined only by  
${d \epsi({\vp})}/{d {\vp}^2} |_{|{\vp}| = p_{F}}$.
T. Suzuki \cite{ISGQR} has given $M^{*}_{L}/M \approx 0.85$ from 
this analysis.
Using this value the enhancement of the space current from the free one 
is about 15 \%.
Namely the enhancement of the magnetic-moment from the Schmidt value is 
also about 15 \%.

When we use the typical value of $M^{*}/M \approx 0.55 - 0.6$ 
\cite{Hama,DBHF,Qing}, 
the enhancement is about 50 \% in the Hartree approximation.
The effect of the momentum-dependence largely improves the magnetic-moment.

Finally we discuss how much degree of momentum-dependence is necessary to get 
$M^{*}_{L}/M = 0.85$ with $M^{*}/M = 0.6$.
For simplicity we neglect the space part of the vector fields, which is
really small in the Hartree-Fock calculation. 
The single particle energy is written as
\begin{equation}
\epsi({\bf p}) = \sqrt{ {\vp}^2 + M^{*2}(\vp) } + U_{0}(\vp).
\end{equation}
The Landau mass is given by
\begin{equation}
M^{*}_{L} = \Pi_{0}(p_{F}) ( 1 - 
M^{*}(p_{F}) \frac{d U_s}{d \vp^2} + 
2 \Pi_{0}(p_{F}) \frac{d U_0}{d \vp^2} )^{-1} .
\end{equation}
In the case that ${d U_s}/{d \vp^2} = 0$, for example,
\begin{equation} 
\frac{U_{0}(p_F) - U_{0}(0)}{U_{0}(p_F)} \approx
 2 p_F^{2} \frac{d U_0}{d \vp^2} / U_0 \approx 0.04 
\end{equation}
with $M^{*}_{L}/M = 0.85$ and $M^{*}/M = 0.6$.
Only 4 \% contribution of the momentum-dependence of the vector-field can reduce
the magnetic-moment about 20 \%.
Hence this argument does not change the saturation property and the nuclear
equation of state as seen in Ref. \cite{TOMO1}.

We can easily imagine that the one-pion exchange force largely produce the
momentum-dependence because the interaction range is largest.
The Fock self-energy from it with the pseudo-scalar coupling gives
${d U_s}/{d \vp^2} > 0$ and ${d U_0}/{d \vp^2} < 0$,
where both contribution can enlarge the Landau mass.
Hence the scenario in this paper is very possible.

In summary we have defined a new current, so called Dirac-Fock current,
which is consistent to the momentum-dependent Dirac fields,
and have shown that the strength of the spacial current is determined by the
Landau mass $M^{*}_L$ independently of the effective mass $M^{*}$.
When we take the value of $M^{*}_{L}$ from the result of ISGQR, 
the enhancement of the magnetic-moment should be 15 \% from 
the Schmidt value even in the isovector case.
In this case the Dirac fields does not need to be varied largely.

In fact the observed magnetic-moment includes the anomalous part,
which is not enhanced by the small effective mass,
and is given by the nucleon on the nuclear surface where the effective
mass is bigger than that at the saturated nuclear matter.
Therefore the actual enhancement of the isovector magnetic-moment 
should be less than 10 \%.
This difference is permissible because the actual isovector magnetic-moment
has other contributions such as the exchange current. 

As seen in this paper the momentum-dependent parts, which are non-local
in the finite nuclei, are very effective even if these parts are small.
In future we need to discuss effects of the non-local parts of Dirac fields
to study nuclear structure and reactions. 

Since the Fock effects are largely seen in the wide energy region 
\cite{KLW1,TOMO1}, this Dirac-Fock current also plays an important role
with the large momentum-transfer $q$. 
In future we need to discuss effects of this current in the high 
momentum transfer phenomena such as the quasielastic electron scattering 
\cite{QES}.
For this discussion we will have to figure out the detailed form of this 
current which has not been given in this paper.
In order to get final conclusions, of course, we must also take into account
the Ring-Diagram \cite{RRPA} which contributes to the final results largely.


\end{document}